\documentclass[12pt]{article}
\usepackage{epsfig}\parskip 5pt plus 1pt
\usepackage{amssymb}
\usepackage{amsmath}
\newcommand{\be}{\begin{equation}}
\newcommand{\ee}{\end{equation}}
\newcommand{\bea}{\begin{eqnarray}}
\newcommand{\eea}{\end{eqnarray}}

\def\simge{\mathrel{%
   \rlap{\raise 0.511ex \hbox{$>$}}{\lower 0.511ex \hbox{$\sim$}}}}
\def\simle{\mathrel{
   \rlap{\raise 0.511ex \hbox{$<$}}{\lower 0.511ex \hbox{$\sim$}}}}

\begin{document}
\thispagestyle{empty}
\vspace*{1cm}
\begin{center}
{\Large{\bf Bandwidth enhancement for parametric amplifiers operated in chirped multi-beam mode} }\\
\vspace{.5cm}
F. Terranova$^{\rm a}$, H. Kiriyama$^{\rm b}$, F. Pegoraro$^{\rm c}$ \\
\vspace*{0.5cm}
$^{\rm a}$ Laboratori Nazionali di Frascati dell'INFN,
Frascati (Rome), Italy \\
$^{\rm b}$ Advanced Photon Research Centre, JAERI, Kizu-cho, Kyoto-fu,  Japan \\
$^{\rm c}$ Dip. di Fisica, Univ. di Pisa and CNISM, Pisa, Italy \\
\end{center}

\vspace{.3cm}
\begin{abstract}
\noindent
In this paper we discuss the bandwidth enhancement that can be
achieved in multi-Joule optical parametric chirped pulse amplification
(OPCPA) systems exploiting the tunability of parametric amplification.
In particular, we consider a pair of single pass amplifiers based on
potassium dideuterium phosphate (DKDP), pumped by the second harmonic
of Nd:glass and tuned to amplify adjacent regions of the signal
spectrum. We demonstrate that a bandwidth enhancement up to 50\% is
possible in two configurations; in the first case, one of the two
amplifiers is operated near its non-collinear broadband limit; to
allow for effective recombination and recompression of the outgoing
signals this configuration requires filtering and phase manipulation
of the spectral tail of the amplified pulses. In the second case,
effective recombination can be achieved simply by spectral filtering:
in this configuration, the optimization of the parameters of the
amplifiers (pulse, crystal orientation and crystal length) does not
follow the recipes of non-collinear OPCPA.
\end{abstract}

\vspace*{\stretch{2}}
\begin{flushleft}
\end{flushleft}

\newpage

\section{Introduction}
\label{introduction}

The use of chirped pulses to amplify high energy signals avoiding
undesired nonlinear effects (CPA) is nowadays a standard technique in
almost any ultrafast laser system~\cite{CPA}. In recent years,
however, interest has increased about the possibility of substituting
the laser amplifier with optical parametric
amplifiers~\cite{review}. This concept, introduced by Dubietis et
al. in 1992~\cite{dubietis} has been proved up to
$\mathcal{O}(10)$~Joule~\cite{collier} both in collinear and
non-collinear mode. Optical parametric chirped pulse amplification
(OPCPA) has several potential advantages with respect to traditional
techniques, the most celebrated being the fact that no energy is
accumulated in the medium except during the amplification time and
both ASE pollution and the overall B-integral of the amplifier can be
substantially reduced. Since this technique is parametric, it exhibits
a high degree of tunability that can be exploited either to reach
narrowband amplification of wavelengths not available on CPA systems
(mainly employing Type II phase matching) or to further increase the
gain bandwidth in Type I amplification.

The simplest method to enhance the OPCPA bandwidth is to add a new
degree of freedom (the angle $\alpha$ between the pump and the signal)
and tune $\alpha$ to reach phase matching at first order for small
deviations from the central signal wavelength (``broadband
non-collinear OPCPA''). In fact, the use of non-collinear beams is a
very well established technique that found experimental confirmation
and a number of applications since the
60's~\cite{noncollinear}. Non-collinear phase
matching~\cite{van_tran,zernike} is ofter implemented in OPCPA and, as
it will be shown, it is a prerequisite for the mode of operation of
the amplifiers discussed in this paper.  On the other hand, tunability
could be exploited in a subtler way to reach ultra-broadband signal
amplification.  A parametric amplifier acts on the seed signal through
a transfer function $H(\Omega)= g(\Omega) \mathrm{e}^{-i
\Gamma(\Omega)}$ that depends on the signal frequency
$\Omega$. Analytical expressions for this function and, particularly,
for the spectral phase $\Gamma(\Omega)$ are discussed in
Sec.\ref{sec:setup}; here we note that the real function $g(\Omega)$
is such that the intensity gain $G(\Omega)$ is greater than 1 only in
a finite domain $[A,B]$, which depends on the angles between the pump,
signal and optical axes of the amplifying crystal.  $\Gamma(\Omega)$
can be expressed as an analytic function and it is bounded to $[-\pi
,\pi]$ in the domain $[A,B]$ ($A,B \in {\mathbb R}$). In the ideal
case, the degree of freedoms available in non-collinear OPCPA could be
tuned to have flat-top\footnote{It means that $g(\Omega) \simeq g$ is
constant in the interval $[A,B]$. \label{flattop}} adjacent
amplification domains:
\be
H(\Omega)= 
\begin{cases}
g_1(\Omega) \mathrm{e}^{-i \Gamma_1(\Omega)} \simeq g \mathrm{e}^{-i
\Gamma_1(\Omega)} & \text{ if $\Omega \in [A,B]$ } \\ 
g_2(\Omega) \mathrm{e}^{-i \Gamma_2(\Omega)} \simeq g \mathrm{e}^{-i
\Gamma_2(\Omega)} & \text{ if $\Omega \in [B=C,D]$ } \\
0 & \text{elsewhere.}
\end{cases}
\ee
In this case, the superposition
of the signals coming from the pair of amplifiers is:
\bea 2\pi  \tilde{E}_{out}(t) =  \int_0^\infty g(\Omega)
\tilde{E}_{in}(\Omega) \mathrm{e}^ {-i \Gamma(\Omega)} \mathrm{e}^{i
\Omega t} \ d \Omega = \nonumber \int_A^B g_1(\Omega) \tilde{E}_{in}(\Omega)
\mathrm{e}^ {-i \Gamma_1(\Omega)} \mathrm{e}^{i \Omega t} \ d \Omega +
\\ 
\mbox{} \int_B^C \left[ g_1(\Omega) \tilde{E}_{in}(\Omega) \mathrm{e}^
{-i \Gamma_1(\Omega)} + g_2(\Omega) \tilde{E}_{in}(\Omega)
\mathrm{e}^ {-i\Gamma_2(\Omega)}  \right] \mathrm{e}^{i \Omega t} \ d \Omega + \nonumber \\
\int_C^D g_2(\Omega) \tilde{E}_{in}(\Omega) \mathrm{e}^ {-i
\Gamma_2(\Omega) } \mathrm{e}^{i \Omega t} \ d \Omega  \simeq 
g\ \int_A^D \tilde{E}_{in}(\Omega)
\mathrm{e}^ {-i \Gamma(\Omega)} \mathrm{e}^{i \Omega t} \ d \Omega
\label{eq:comb}
\eea
providing an effective bandwidth $[A,D]=[A,B]+[B,D]$. In this formula
$\Gamma(\Omega)$ is $\Gamma_1 \cup \Gamma_2$ over the domain $[A,D]$,
$\tilde{E}_{in}(\Omega)$ is the signal spectral amplitude and
$\tilde{E}_{out}(t)$ is the complex electric field of the amplified
signal in the time domain. Note that in Eq.\ref{eq:comb}, the phase
relations between the pump and the signal do not appear, while the
phase of the amplified Fourier-component is given only by the function
$\Gamma(\Omega)$.  This is a general feature of parametric
amplification, in which the phase difference between pump and signal
is transferred to the idler, thus compensating for random differences
between the two. In all practical implementations, this requires the
removal of the idler between the two amplification domains.  The
concept of superposition of adjacent amplification domains has been
studied in the framework of two-beam pumping optical parametric
amplification~\cite{zeromskis}. In fact, multiple pump beams have been
first exploited by Smilgevicius and Stabinis for spatial bandwidth
reduction~\cite{smilgevicius} and, later on, combination of multiple
incoherent pumps has been achieved by Marcinkevicius et
al.~\cite{marcinkevicius}. Again, possible scaling to high powers in the
framework of OPCPA has been discussed numerically in~\cite{wang}.

On the other hand, a practical realization suited for high energy
pulses that could be employed, e.g., for proton and electron
acceleration at high repetition rates in nuclear and particle physics
applications~\cite{terranova} has to face several additional
difficulties.  First of all, the need of amplification in the
multi-Joule regime reduces substantially the choice of non-linear
crystals that can be employed (mainly KDP and DKDP) and, therefore,
the range of adjacent domains available in practice.  Moreover, the
actual behaviour of non-linear crystals does not fulfill the condition
of ideal flat top response in the sense of Footnote \ref{flattop}.  If
adjacent amplification is obtained pumping simultaneously the same
crystal with two pumps and a single seed there is no way to filter
unwanted amplified Fourier components. Moreover, to allow for
effective recompression, the phase of amplified signal should be a
regular function of $\Omega$ around the overlap region $[B,C]$.  We
demonstrate in Sec.\ref{sec:results} that this requirement is
incompatible with the requirement $[A,D] \gg [A,B]$.  This is the
reason for large phase and gain fluctuations predicted in multi-beam
OPCPA before recompression.  Much larger flexibility in pulse cleaning
and phase adjustment~\cite{witte,tavella} is allowed by two amplifiers
seeded from the same split broadband signal; this comes, however, at
the price of signal recombination before recompression.  In the
following, we consider specifically the single pass multi-seed
bandwidth enhancement for a setup suited for high energy pulse
amplification and based on DKDP (Sec.\ref{sec:setup}). We show that a
bandwidth enhancement up to 50\% is actually possible in two
configurations (Sec.\ref{sec:results}): the first one requires
filtering and phase manipulation of the amplified signal in the
neighboring region $[B,C]$ and operates one of the two amplifiers near
its maximum bandwidth. Having one of the two amplifiers near its
single beam broadband condition (see Eq.\ref{eq:maxb} below) is a
standard choice in literature. Still this choice is not mandatory: a
complete multi-dimensional optimization (Sec.\ref{sec:results})
indicates that in dedicated regions of the parameter space smooth
spectral responses can be achieved without compromising the bandwidth
enhancement. Here, the regularization of the response corresponds to
simple frequency filtering and an overall spectral phase advance of
one of the two signals.  Clearly, in this second configuration, the
optimization of the parameters of the amplifiers does not follow the
recipes of non-collinear OPCPA.

\section{Multi-beam operation of DKDP amplifiers}
\label{sec:setup}

\begin{figure}
\centering \includegraphics[width=12cm]{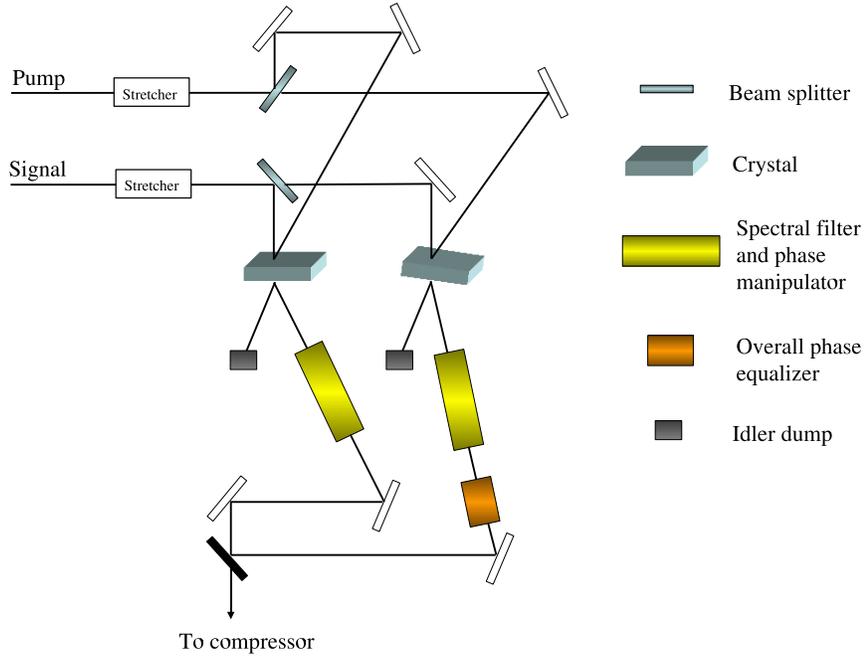}
\caption{Schematics of the multi-beam OPCPA setup (angles are not in scale).}
\label{fig:scheme}
\end{figure}

Multi-Joule amplifiers for Inertial Confined Fusion and particle
acceleration will likely be pumped by harmonics of solid state
lasers~\cite{terranova}.  Present state of the art amplifiers are
based on the second harmonic of Nd:glass (527 nm) and future,
high-repetition rate systems could be based on high efficiency
diode-pumped solid-state lasers. Several nonlinear crystals can
exploit these wavelengths (LBO, BBO etc.) but presently only KDP and
its isomorph DKDP can be grown to apertures of 30~cm or more.  In
order to get amplified signals in the near-IR, where high power
recompression optics is more readily available, KDP can be operated in
quasi degenerate mode while it has been demonstrated that DKDP,
operated in non-collinear non-degenerate mode, provides a significantly
larger bandwidth~\cite{lozhkarev}. In the following, we consider as
testbed for multi-beam operation a set of amplifiers based on DKDP
and pumped by the second harmonic of Nd:YLF (527 nm) stretched up to
$\sim$ 0.5~ns (Fig.\ref{fig:scheme}). 

\begin{figure}
\centering \includegraphics[width=8cm]{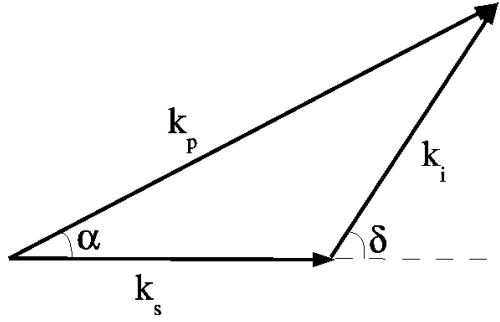}
\caption{Phase matching triangle for non-collinear OPCPA.}
\label{fig:vectors}
\end{figure}

The DKDP amplifiers can be operated in non-collinear mode.  In this
case the pump and signal wave vectors $k_p$ and $k_s$ form an angle
$\alpha$ between them (Fig.\ref{fig:vectors}). The angle is
independent of the signal wavelength and the idler frequency is fixed
by energy conservation ($\omega_p = \omega_s + \omega_i$) but the
emission angle $\delta$ varies with the signal wavelength
$\lambda_s$. Phase matching is achieved when:
\begin{eqnarray}
\Delta k_{||} & = &  k_p \cos \alpha - k_s -k_i \cos \delta = 0 \nonumber \\ 
\Delta k_{\perp} & = &  k_p \sin \alpha -k_i \sin \delta = 0  \nonumber
\end{eqnarray}

\noindent
The additional degree of freedom coming from the introduction of
$\alpha$ can be exploited to improve the gain bandwidth. In
particular, it exists an $\alpha$ such that phase matching is achieved
at first order for small deviations from the central signal wavelength
(``single beam broadband condition'').  It can be
demonstrated~\cite{cerullo} that this condition corresponds to choosing
\be \sin \alpha = \frac {k_i}{k_p} \sin ( \mathrm{acos} \left[
n_{gi}/n_{gs} \right] ) \label{eq:maxb} \ee
where $n_{gs} = c \mathrm{d} k_s/ \mathrm{d} \omega_s$ and $n_{gs} = c
\mathrm{d} k_i/ \mathrm{d} \omega_i$.  The derivatives can be computed
from the Sellmeier's equation\footnote{In the following we use $$
n_0^2 =2.2409 + \frac{0.0097}{\lambda^2 - 2.2470 } + 0.0156 \
\frac{\lambda^2}{\lambda^2- 126.9205} $$ for the ordinary index and
$$ n_e^2 = 2.1260 + \frac{0.0086}{\lambda^2 - 0.7844} + 0.0120 \
\frac{\lambda^2}{\lambda^2-123.4032} $$ for the principal
extraordinary index ($\lambda$ is the wavelength in $\mu$m).}  for
DKDP.  Note, however, that the single-beam broadband condition is not
necessarily the best condition also for multi-beam amplifiers and, in
general, we will not request $\alpha$ to fulfill the constraint of
Eq.\ref{eq:maxb}.

Analytic expressions for the intensity gain $G=G(\Omega)$
and the phase of the amplified signal $\Gamma(\Omega)$ are available
solving the coupled wave equations for signal, idler and pump in the
slowly varying envelope approximation and assuming no pump
depletion~\cite{ross,armstrong,ross_opt_soc}. In this case,
\be G \ = \ 1+(\gamma L)^2 \ \left[ \frac{\mathrm{sinh} B}{B} \right]^2 \ee
where $B\equiv \left[ (\gamma L)^2 - (\Delta k L/2)^2 \right]^{1/2}$;
$\gamma$ represents the gain coefficient
\be
\gamma \equiv 4 \pi d_{eff} \sqrt{ \frac{ I_p }{2 \epsilon_0 
\ n_{ep}(\theta_m) \ n_{os} \ n_{oi} \ c \ \lambda_s \ \lambda_i } };
\ee 
\noindent while the quantity $L$ is the length of the crystal and
$\Delta k \equiv k_p-k_s-k_i$ is the phase mismatch among signal,
idler and pump.  $d_{eff}$ is the effective nonlinear coefficient for
Type I phase matching~\cite{boyd} in DKDP. 
In Sec.\ref{sec:results}, as well as in Ref.~\cite{ross}, $L$ has been
equalized to reach $G=1000$ and, for the present setup, $L\sim$ 1~cm.

\noindent
Within the same approximations, the spectral phase $\Gamma(\Omega)$ is
\begin{equation}
 \mathrm{arctan} \left[ \frac{ B \sin{A} \ \mathrm{cosh} B-A \cos{B} \ \mathrm{sinh} B  }
{ B \cos A \ \mathrm{cosh} B + A \sin A \ \mathrm{sinh} B } \right]
\label{equ:phase}
\end{equation}
$2A$ being the product of the phase mismatch and the crystal length.

Fig.\ref{fig:scheme} shows a scheme of the amplifier pair seeded by
the same broadband signal, which is split before reaching the two
crystals. The seed signal is split and sent to the two amplifiers at
different angles $\theta_1$ and $\theta_2$ with respect to the optic
axis of the crystals, achieving phase matching for different central
frequencies.  Similarly, the split pumps impinge upon the nonlinear
crystals at an angle $\alpha_1 \neq \alpha_2$ with respect to the
seed. As usual in OPCPA, strong constraints are put on the
synchronization of the pump and beam pulse while the synchronization
of the two chirped ($\tau \simeq$ 0.5~ns) signals coming from the same
source and crossing only passive optical elements do not add
additional difficulties except for the equalization of the overall
optical path after filtering and phase manipulation (yellow box in
Fig.\ref{fig:scheme}). In principle, more than two amplifiers could be
operated in parallel within the parameter range in which phase
matching is possible. Single-beam broadband conditions for DKDP can be
fulfilled in the range 790-1050~nm~\cite{lozhkarev}, hence we expect
maximum bandwidths greater than 3100~cm$^{-1}$.

\section{Numerical results}
\label{sec:results}

\begin{figure}
\centering \includegraphics[width=12cm]{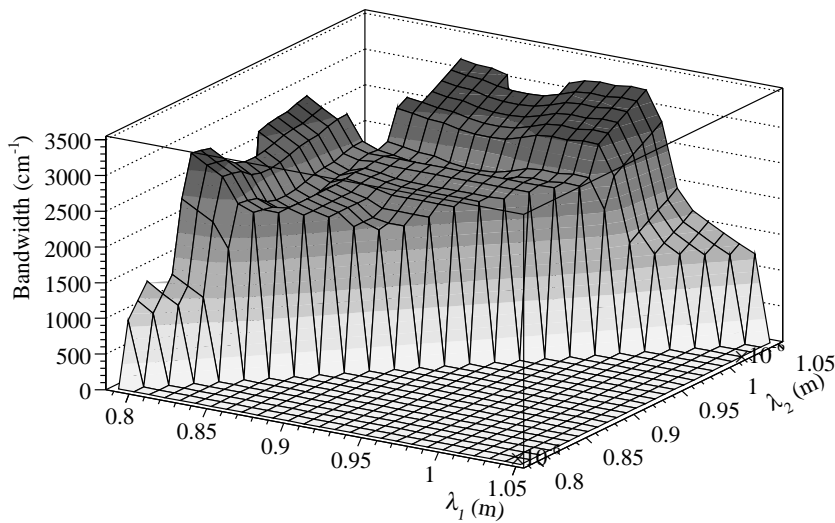}
\caption{Effective bandwidth in cm$^{-1}$ as a function of the central
wavelength $\lambda_1$ and $\lambda_2$ on which the two parametric
amplifiers have been tuned. Only the region $\lambda_1 < \lambda_2$ is
plotted.}
\label{fig:runbandw}
\end{figure}

As mentioned above, the single-beam broadband condition provides only
a first estimate for the tuning of the parameters of the
amplifiers. These are the crystal lengths $L_1$ and $L_2$, the signal
to optical axis angles $\theta_1$ and $\theta_2$ and the signal to
pump angles $\alpha_1$ and $\alpha_2$. Once $\theta_{1,2}$ and
$\alpha_{1,2}$ are fixed, $L_1$ and $L_2$ can be computed numerically
to have fixed maximum gain for each amplifiers ($10^{3}$ in the
present case). Therefore, in the occurrence of a flat top response,
$|E_{out}(\Omega)|^2$ remains constant within the effective
amplification domain $[A,D]$. In this section, we investigated
numerically the region $\lambda_1,\lambda_2 \in [790,1050]$~nm. For
each point of the grid, $\alpha$ is tuned around the single beam
broadband condition. We impose however, as an additional constraint,
the absence of local amplification maxima in order to identify the
parameters where a nearly flat-top response occurs.  The effective
bandwidth $[A,D]$ assuming perfect filtering is shown in
Fig.~\ref{fig:runbandw}. Note that if the intensity gain around the
overlap region $[B,C]$ falls below its half-maximum, the signals are
considered unsuitable for multi-beam operation and the effective
bandwidth is simply the one of the broader amplifier. Hence, in the
region $\lambda_1\simeq \lambda_2 \simeq$~910~nm, the effective
bandwidth approaches the corresponding value for ultrabroadband phase
matching in DKDP~\cite{lozhkarev} ($\sim 2000$~cm$^{-1}$).  Note also
that the grid is symmetric for the permutation $\lambda_1
\leftrightarrow \lambda_2$ and, for sake of clarity, only the
$\lambda_1< \lambda_2$ region is plotted in Fig.~\ref{fig:runbandw}.

\begin{figure}
\centering \includegraphics[width=10cm]{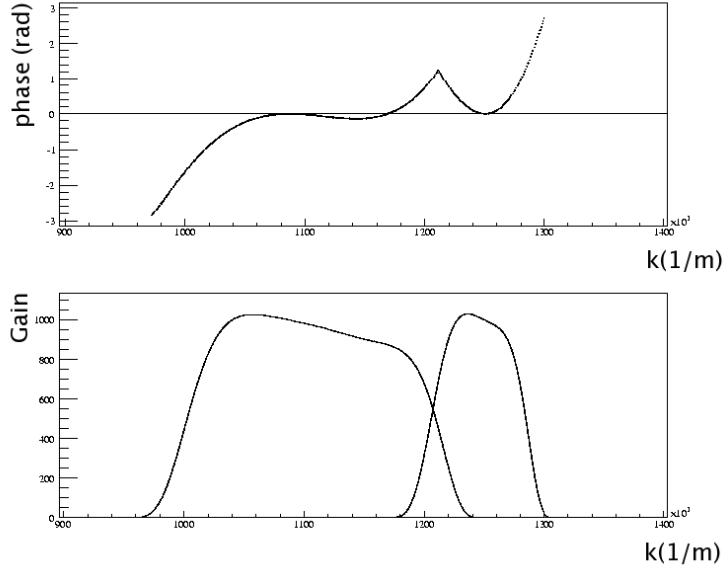}
\caption{Spectral phase (upper plot) and intensity gain (lower plot)
versus wavenumber for two amplified signals with central wavelength
$\lambda_1= 920$~nm and $\lambda_1= 800$~nm.}
\label{fig:twobeam_800_920}
\end{figure}

As anticipated, effective bandwidths exceeding 3000~cm$^{-1}$ are
possible already with two amplifiers if one of the two is operated in
the vicinity of its single-beam broadband condition (Eq.\ref{eq:maxb})
and the other is tuned to have adjacent flat-top response.  In order
to maximize $[A,D]$, however, the central frequency of the second
amplifier must be significantly distant from the overlap
region $[B,C]$ and, in such region, we expect $\Gamma(\Omega) \ne
0$. Moreover, $|d\Gamma / d\Omega|$ increases monotonically far from
the central frequency $\lambda_{1,2}$. In the proximity of the
ultrabroadband phase matching condition\footnote{It corresponds to
$\lambda_1 \simeq 910$~nm with $\alpha$ fulfilling the constraint of
Eq.\ref{eq:maxb}.}, $\Gamma(\Omega)$ has always an inflection around
the central frequency $\Omega=2\pi c/\lambda_{1,2}$ while far from
this region ($\lambda \gg $ or $\ll$ 910~nm) it has a
global minimum ($\lambda \ll $~910~nm) or maximum ($\lambda \gg
$~910~nm). This situation is depicted in Fig.\ref{fig:twobeam_800_920}
for $\lambda_1=920$~nm and $\lambda_2=800$~nm. The top plot represents
the spectral phase $\Gamma$ as a function of the wavenumber for the
two amplified signals. The corresponding intensity gain is shown in
the lower plot. This condition of strong spectral phase mismatch near
the overlap region is very general and it is the result of operating
one of the two amplifiers near the ultrabroadband condition and
requiring a substantial bandwidth increase from the second.  It
results in strong phase fluctuations and, as already mentioned, its
exploitation requires phase manipulation and filtering of the spectral
region between $\lambda_1$ and $\lambda_2$.

\begin{figure}
\centering \includegraphics[width=\textwidth]{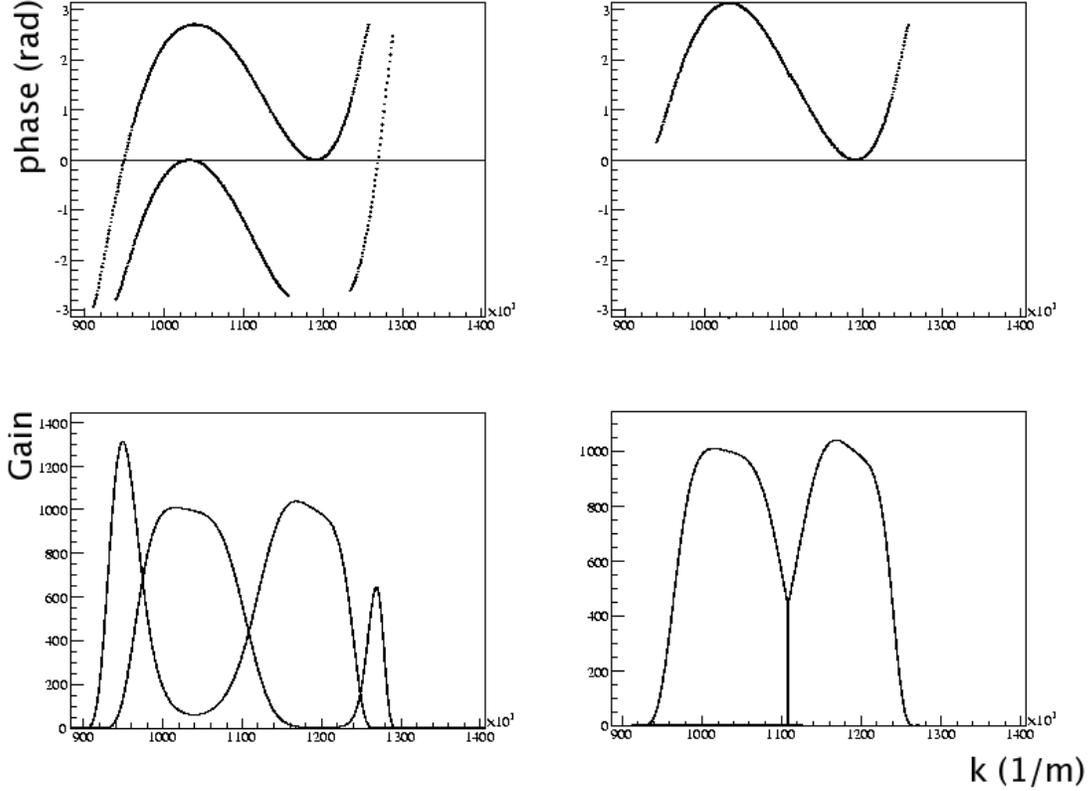}
\caption{Spectral phase (upper plot) and intensity gain (lower plot)
versus wavenumber for two amplified signals with central wavelength
$\lambda_1= 970$~nm and $\lambda_1= 840$ and re-optimization of the angles 
between signals and pumps. The left plots are unfiltered and the right ones 
show  $G$ and $\Gamma$ after filtering of the disconnected satellites and phase
advancing ($+\pi$) of the signal at $\lambda_1=970$~nm.}
\label{fig:twobeam_840_970_noalphatune}
\end{figure}

If both the amplifiers are operated far from the ultrabroadband
parameter range, inflections do not occur and the overall phase
mismatch between the two signals is just $\pi$. The spectral phase and
intensity gain are shown in Fig.\ref{fig:twobeam_840_970_noalphatune}
for $\lambda_1=970$~nm, $\lambda_2=840$~nm and $\alpha$ re-optimized
to have a smooth spectral mismatch in the overlap region.  In this
case each amplification domain has a disconnected satellite that has
to be filtered after amplification. $G$ and $\Gamma$ are plotted
before the filtering in the left plots of
Fig.\ref{fig:twobeam_840_970_noalphatune}. The right plots show the
corresponding functions after filtering and applying an overall phase
shift of $\pi$ to the $\lambda_1=970$~nm signal.  Clearly, in this
case, no spectral phase manipulations are needed to get a smooth phase
response. This simplifies remarkably the design of the
recompressor. Still, the design and engineering of the recompressor,
together with the corresponding overall conversion efficiency, are not
considered in the present work and deserves a dedicated study.

\section{Conclusions}

The high degree of tunability of parametric amplification can be
exploited in several manners even in the multi-Joule energy range.
While Type II amplification is mainly used to get narrowband signals
over a broad tuning ranges, in Type I tunability can be exploited to
increase remarkably the gain bandwidth. Tunability allows the
construction of multiple systems with adjacent amplification domains
so that the overall gain domain exceed substantially the one of a
single amplifier.  In this paper we considered specifically a setup
aimed at the amplification of high energy pulses in the near-IR and
based on potassium-dideuterium-phosphate (DKDP). Faint broadband
signals are chirped, split and sent to several DKDP amplifiers; the
latter have been tuned to have adjacent amplification domains, while
the amplified signals are recombined before injection into the
compressor.  Numerical simulations indicates that the effective
bandwidth can be enhanced up to 50\% in two specific configurations
(Sec.\ref{sec:results}): the first one requires filtering and phase
manipulation of the neighborings regions and operates one of the
two amplifiers at its own  maximum bandwidth. The other
results from a dedicated optimization and needs only filtering after
amplification: in this case the tuning of the parameters of the
amplifiers does not follow the recipes of single-beam non-collinear
OPCPA (Eq.~\ref{eq:maxb}).

\section*{Acknowledgments}
We wish to thank J.~Collier for bringing our attention to
DKDP-based systems. We also thank S.~Bulanov and P.~Migliozzi for useful
discussions on the applications of multi-J OPCPA to nuclear and
particle physics.


\end{document}